\documentclass[a4paper]{article}

\usepackage{a4wide}
\usepackage[round]{natbib}
\usepackage{amsmath}
\usepackage[latin1]{inputenc}
\usepackage{graphicx}

\makeatletter
\catcode`\"=\active
\def"#1{{\ensuremath{\csname gr@sk#1\endcsname}}}

\let\gr@ska=\alpha
\let\gr@skb=\beta
\let\gr@skc=\chi
\let\gr@skd=\delta
\let\gr@ske=\epsilon
\let\gr@skf=\phi
\let\gr@skg=\gamma
\let\gr@skh=\eta
\let\gr@ski=\iota
\let\gr@skj=\varphi
\let\gr@skk=\kappa
\let\gr@skl=\lambda
\let\gr@skm=\mu
\let\gr@skn=\nu
\let\gr@sko=o
\let\gr@skp=\pi
\let\gr@skq=\theta
\let\gr@skr=\rho
\let\gr@sks=\sigma
\let\gr@skt=\tau
\let\gr@sku=\upsilon
\let\gr@skv=\varomega
\let\gr@skw=\omega
\let\gr@skx=\xi
\let\gr@sky=\psi
\let\gr@skz=\zeta

\let\gr@skA=A
\let\gr@skB=B
\let\gr@skC=X
\let\gr@skD=\Delta
\let\gr@skE=\varepsilon
\let\gr@skF=\Phi
\let\gr@skG=\Gamma
\let\gr@skH=H
\let\gr@skI=I
\let\gr@skJ=\vartheta
\let\gr@skK=K
\let\gr@skL=\Lambda
\let\gr@skM=M
\let\gr@skN=N
\let\gr@skO=O
\let\gr@skP=\Pi
\let\gr@skQ=\Theta
\let\gr@skR=P
\let\gr@skS=\Sigma
\let\gr@skT=T
\let\gr@skU=Y
\let\gr@skV=\varsigma
\let\gr@skW=\Omega
\let\gr@skX=\Xi
\let\gr@skY=\Psi
\let\gr@skZ=Z

\makeatother

\def\egcite#1{\citep[e.g.][]{#1}}

\let\vec\boldsymbol				
\let\l=\left \let\rr=\right
\def\k{\ensuremath{\vec k}}
\def\v{\ensuremath{\vec v}}

\def\sq#1#2{\ensuremath{#1\,\mathrm{#2}}}

\title{The Capabilities of the EISCAT Svalbard Radar
    for Inter-hemispheric Coordinated Studies}

\author{Tom Grydeland\thanks{Dept.~of Physics, University of Tromsø, N-9037 Tromsø, Norway}
\and    Anja Strømme\footnotemark[1]
\and    Tony van Eyken\thanks{EISCAT Scientific Association,
Currently at SRI International, Menlo Park, California 94025, USA}
\and    Cesar La Hoz\footnotemark[1]}

\begin{document}

\maketitle

\begin{abstract}
    In this article we want to present the EISCAT Svalbard Radar (ESR)
    in some detail, as well as some of the instruments of interest for
    ionospheric and magnetospheric research that are located in the
    vicinity of it.  We particularly describe how this instrument cluster,
    close to the geomagnetic conjugate point of the Chinese Antarctic
    Zhongshan station, can contribute to inter-hemispheric coordinated
    studies of the polar Ionosphere.

    Keywords: EISCAT, Incoherent Scatter Radar, Conjugate studies
\end{abstract}


\section{The Incoherent Scatter Technique}

The use of incoherent scatter radars as a powerful ground-based diagnostic
tool for studying the near-Earth space environment began with the first
theoretical predictions by \citet{Gordon:1958}, and the first observations
by \citet{Bowles:1958} a few months later.  There are several
comprehensive reviews of the Incoherent Scatter technique,
\egcite{Evans:1969,Evans:1975,Bauer:1975,Beynon:1978}, while an
overview of the early history of Incoherent Scatter, as well as an updated
description of the theory, instruments and signal processing involved can
be found in \citep{KRS:Farley}.

The term \emph{Incoherent Scatter} (IS) from an ionized gas refers to the
extremely weak scatter from fluctuations in plasma density caused by the
random thermal motion of the ions and electrons.
Due to the very low radar scattering cross section of an individual
electron, only about \sq{1.0\times 10^{-28}}{m^{2}}, the total cross
section of all the electrons in ten cubic kilometers --- a typical volume
probed in an experiment --- of the ionosphere with
a maximum density of the order of \sq{10^{12}}{m^{-3}} is only about
\sq{10^{-6}}{m^{2}}. To detect signal from this weak source, we need a
radar capable of detecting a coin at \sq{300}{km} distance!

Incoherent Scatter Radars (ISR) therefore consist of large antennas, powerful
transmitters, and sensitive and sophisticated receiver systems, since in
addition to measuring the signal and its power we also need to measure the
full Doppler power spectrum or equivalently the auto correlation function
(ACF) of the back-scattered signal.

When the radar frequency, $"w_\mathrm{radar}$ is much higher than the
plasma frequency $"w_{p}$, the radar wave travels almost unperturbed
through the very dilute ionospheric plasma. A small fraction of the wave
energy is deposited into the acceleration of the electrons, which radiates
back as small dipoles. The ions also absorb energy in this process, but due
to their relatively high mass, their radar scattering cross section is a
factor $(m_{e}/m_{i})^2$  smaller than the electron radar scattering cross
section, and their contribution to the scattered signal is negligible. 

In the pioneering work by Gordon, he assumed that one would see ``true''
incoherent scattering from the individual free electrons. This would
result in a back-scattered power spectrum with a width proportional to the
electron thermal velocity. The first experiments, however, showed a power
spectrum with a width proportional to the ion thermal velocity, with the
power contained in a much narrower frequency range than predicted. This
dramatically improved the signal to noise ratio within this frequency
range. Despite the fact that the electrons are the scattering particles in
this process, they are, due to their low mass, highly mobile and will
therefore easily follow the heavier ions through electromagnetic
interactions.  The typical scale size of these interactions is called the
\emph{Debye length}.  For a typical incoherent scatter radar
configuration, the radar wavelength is much larger than the Debye length
of the ionospheric plasma, and the power spectrum of the received signal
carries information about the plasma as a whole, with the electron
dynamics strongly influenced by the ions.  For the situation with the
radar wavelength shorter
than the Debye length, we have ``true'' incoherent scattering, and the
electrons scatter as free particles, with a received power spectrum
typical for the electron velocity distribution function.

The scattered signal contains information about the plasma density, the
electron and ion temperatures, the ion composition, the bulk plasma motion,
and various other parameters and properties of the probed plasma.

A number of different approaches all lead to the same result for the
incoherent scatter power spectral shape \citep{Fejer:1960, Dougherty:1960,
Salpeter:1960, Rosenbluth:1962, Rostoker:1964, Trulsen:1975}, and this
back-scattered power spectrum can be given by the equation:
\begin{equation}
\begin{split}
    S(\k,"w)={}&N_{e}\l | 1- \frac{"c_{e}(\k,"w)}{"e(\k,"w)}\rr |^{2}\int d\v
    f_{e}(\v)"d("w -\k\cdot\v)\\
    &{}+\sum_{i} N_{i}\l|
    \frac{"c_{e}(\k,"w)}{"e(\k,"w)}\rr|^{2}\int d\v f_{i}"d("w-\k\cdot\v)
\end{split}
\label{eq:powerspectra}
\end{equation}
where the electric susceptibility $"c_{"a}(\k,"w)$ for species "{a} is given by
\begin{equation}
    "c_{"a}(\k,"w)=\frac{"w_{pe}^{2}}{k^{2}}\int_{L}
	\frac{\k\cdot "d_{\v}f_{"a}(\v)} {"w-\k\cdot\v}\,d\v,
\end{equation}
the dielectric constant function $"e(\k,"w)$ is given by
\begin{equation}
    "e(\k,"w)=1+"S_{"a}"c_{"a}(\k,"w),
\end{equation}
and $f_{"a}(\v)$ the velocity distribution function for the species $"a$.

\begin{figure}[p]
  \begin{minipage}[b]{0.5\linewidth}
    \centering\includegraphics[width=2.6in,height=2.6in]%
       {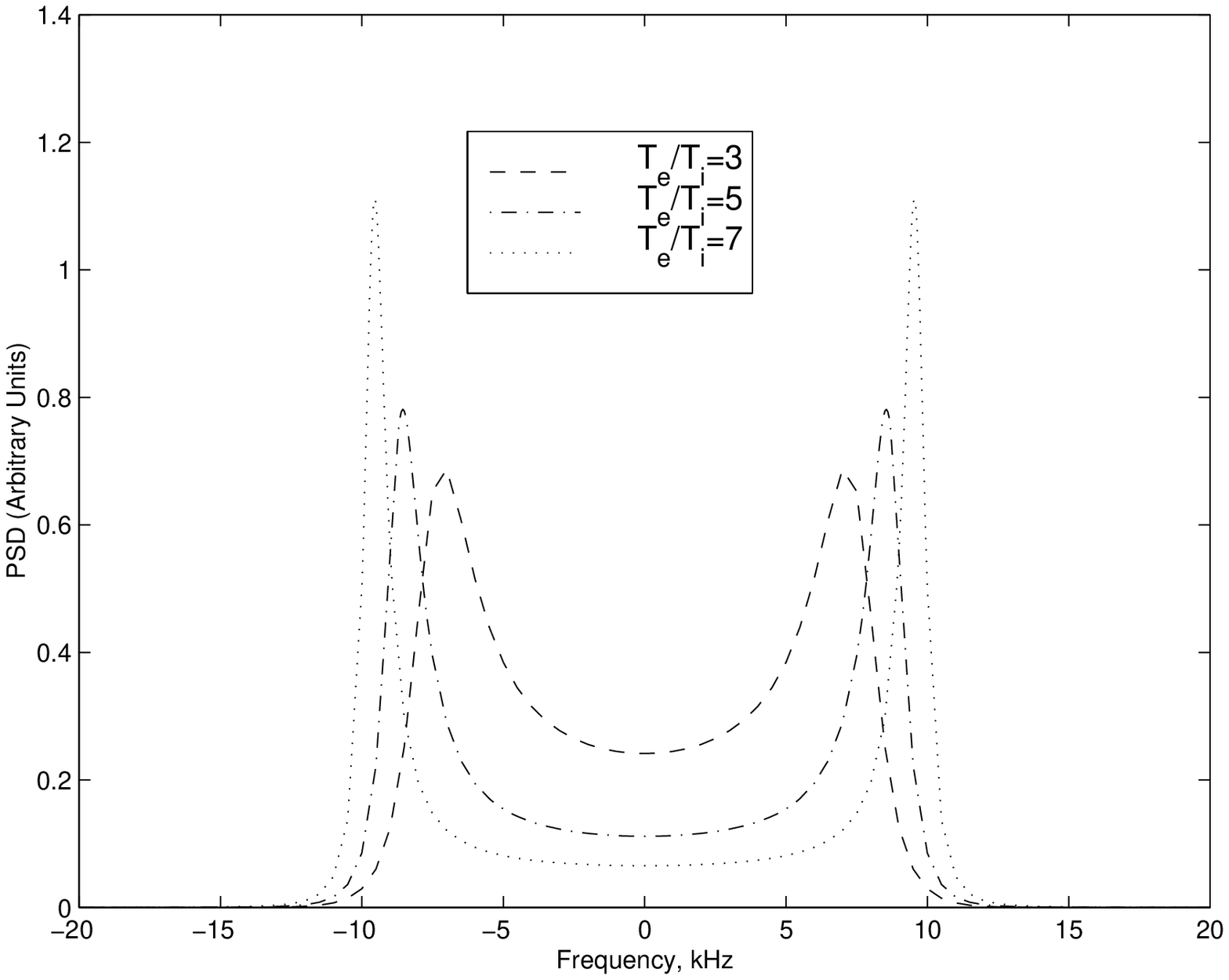}
  \end{minipage}
  \begin{minipage}[b]{0.5\linewidth}
    \centering\includegraphics[width=2.6in,height=2.6in]%
       {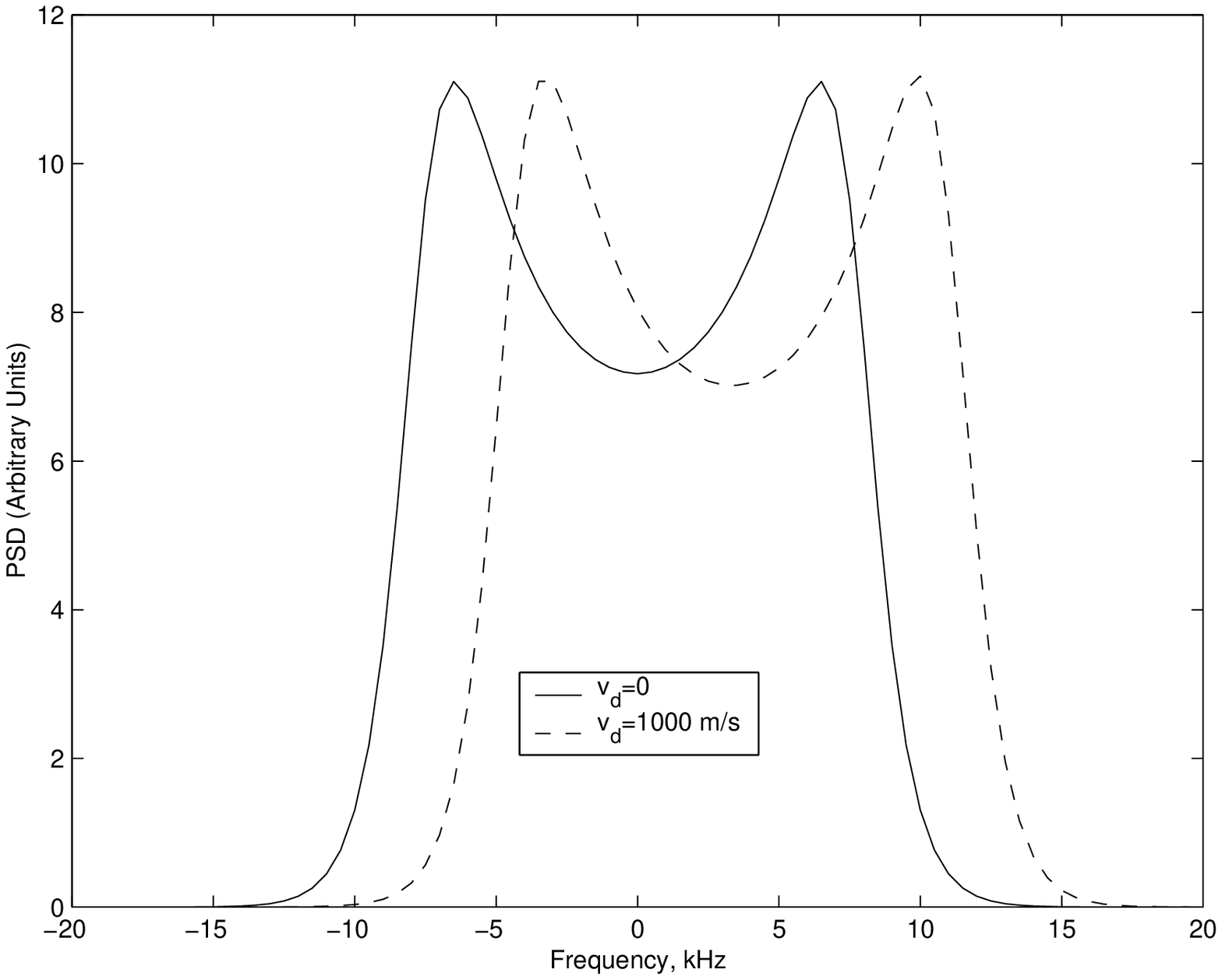}
  \end{minipage}
\caption{Figure a) shows the $T_{e}/T_{i}$ dependence on the ion spectra,
and figure b) shows the effect of plasma drift.}
  \label{fig:tetidrift}
\end{figure}

Equation~\eqref{eq:powerspectra} gives rise to two parts of the power
spectrum, the \emph{ion line} (from the second term) and the \emph{plasma
line} (from the first term) respectively. The portions of the power
spectrum with small Doppler shifts is often referred to as the \emph{ion
line}. It can be viewed as two very broadened overlapping lines
corresponding to damped ion-acoustic waves traveling parallel and
anti-parallel to the $\k$-vector determined by the radar system; toward
and away from the radar for a mono-static system.  The lines have Doppler
shifts corresponding to the frequency of ion-acoustic waves, which are
solutions of equation~\eqref{eq:powerspectra} in the low frequency
range. From the ion-line we are able to determine a range of plasma
parameters:

\noindent\textbf{The electron density, $N_{e}$}
    can be found from the total back-scattered power. The constant of
    proportionality between the electron density and the back-scattered
    power depends on the electron and ion temperature ($T_{e}$ and
    $T_{i}$).\\
\textbf{The temperature ratio, $T_{e}/T_{i}$}
    can be determined from the ratio of the peaks to the dip in the ion
    spectra (shown in figure~\ref{fig:tetidrift}, panel 1) due to the
    $T_{e}/T_{i}$ dependence in the ion-acoustic damping term.\\
\textbf{The ion temperature to mass ratio, $T_{i}/m_{i}$}
    can be found from the width of the ion spectra. If $m_{i}$ is known
    (e.g. from a model), $T_{i}$ can be found, and hence $T_{e}$.\\
\textbf{Line-of-sight ion velocity}
    can be determined from the Doppler shift of the ion spectra (shown in
    figure~\ref{fig:tetidrift}, panel 2). By using a tristatic radar, the
    drift can be determined in three directions, and hence the full ion
    velocity vector can be found. 

The analysis is done by fitting all parameters simultaneously in an
iterative process. Behind this fitting routine, the most severe
assumptions are the homogeneity and stationarity assumed over the whole
scattering volume and the whole integration time. For quiet conditions
these assumptions are sufficiently fulfilled, but during active and
disturbed periods, the returned power is sometimes increased by one or two
order of magnitude, the plasma is driven out of thermal equilibrium,
and the ion-line can be strongly asymmetric with one or both of the
ion-acoustic shoulder enhanced \cite[and references
therein]{Sedgemore-Schulthess:2001}. During these periods, the
fitting process does not work, since it assumes the plasma to be in
thermal equilibrium, and we are not able to determine the plasma
parameters from the spectra.  The processes behind these ``anomalous'' ion
spectra are not yet fully understood, and more work has to be done in
order to understand them, before we will be able to to analyse data
also from these periods.

For much higher frequencies, two narrow less damped lines, the
\emph{plasma lines} are found, one up- and one down-shifted. They are high
frequency solutions of equation \eqref{eq:powerspectra}. Their frequency
depends directly on the electron density, with a small correction from
electron temperature, and they can therefore be used to determine these
plasma parameters when measured.


\section{EISCAT and the ESR}

EISCAT was founded by the six European countries France, Great Britain,
Germany, Norway, Finland and Sweden in 1975 for the purpose of
constructing an ISR at high latitude; right underneath the auroral zone.
The EISCAT UHF radar --- situated outside Tromsø, Norway --- was
inaugurated in 1981.  It is a tristatic system, the only
one in operation in the world today, meaning that in addition to the
transmitter and receiver in Tromsø, there are passive receivers on two
other sites: Sodankylä in Finland and Kiruna in Sweden.  The three sites
give the radar its unique capability of obtaining true vector velocities
in a single common volume.  The EISCAT VHF radar, co-located with the UHF
radar, became operational in 1985, and extended the capabilities of the
EISCAT system in the extreme low- and high-altitude regimes.  The mainland
EISCAT radars were described in \citep{Folkestad:1983}, but the UHF
transmitters and all receiver systems have since been totally redesigned,
so this description is now out of date.  A summary of the EISCAT mainland
radars' capabilities as of the summer of 2001 is given in
table~\ref{eiscattable}.

The EISCAT Svalbard Radar (ESR) was inaugurated in August 1996, the same
year that Japan became the seventh member of EISCAT.  It is
situated on top of Mine 7 on the Breinosa mountain \sq{12}{km} outside
Longyearbyen, the biggest settlement in the Svalbard archipelago.  The ESR
System is described in detail in \citep{Wannberg:1997}.  Since then, the
transmitter has been upgraded to \sq{1}{MW} peak power, and a second
antenna, \sq{42}{m} in diameter and fixed to point along the geomagnetic
field, has been added.  A summary of the ESR parameters is given in
table~\ref{esrtable}.


\begin{table}[p]
\begin{tabular}{lllllll}

Location  		& \multicolumn{2}{c}{Tromsø}& Kiruna	& Sodankylä      	\\ \hline
Geograph. Coordinates	& \multicolumn{2}{c}{69°35´N}	& 67°52´N	& 67°22´N	\\
  		 	& \multicolumn{2}{c}{19°14´E}	& 20°26´E	& 26°38´E	\\
Geomagn. Inclination	& \multicolumn{2}{c}{77°30´N}	& 76°48´N	& 76°43´N	\\
Invariant Latitude	& \multicolumn{2}{c}{66°12´N}	& 64°27´N	& 63°34´N	\\
Band	                & VHF		& UHF		& UHF		& UHF		\\ \cline{2-5}
Frequency (MHz)         & 224		& 929		& 929		& 929		\\
Max. TX bandwith (MHz)  & 3		& 4		& -		& -		\\
Transmitter 	        & 2 klystrons	& 2 klystrons	& -		& -		\\
TX Channels	        & 8 		& 8		& -		& -		\\
Peak power (MW)         & $2\times1.5$	& 2.0		& -		& -		\\
Average power (MW)      & $2\times0.19$ & 0.3		& - 		& -		\\
Pulse duration (msec)	& .001--2.0	& .001--2.0	& -		& -		\\
Phase coding		& binary 	& binary	& binary	& binary	\\
Min. interpulse (msec)	& 1.0		& 1.0		& -		& -		\\
System temp. (K)	& 250-350      & 70-80		& 30-35		& 30-35		\\
Receiver		& analog-digital & \multicolumn{3}{c}{analog-digital}		\\
Digital processing	& \multicolumn{4}{c}{14 bit ADC,}				\\
		        & \multicolumn{4}{c}{Lag profiles 32 bit complex}		\\
			&		&		&		&  	    	\\
Antenna	     & cylinder	                & dish		& dish		& dish		\\
	     & $\sq{120}m\times\sq{40}m$ & \sq{32}m	& \sq{32}m	& \sq{32}m	\\
Feed system  & line feed,		& Cassegrain    & Cassegrain	& Cassegrain	\\
	     & \multicolumn{2}{l}{128 crossed dipoles}	\\
Gain (dBi)   & 46			& 48		& 48		& 48		\\
Polarization & circular			& circular	& any		& any		\\

\end{tabular}

\caption{Table summarising the EISCAT mainland radars} \label{eiscattable}
\end{table}

\begin{table}[p]
\begin{tabular}{lllllll}

Location  		& \hspace{1.0cm} & \multicolumn{2}{l}{Longyearbyen} \\ \hline 
Geograph. Coordinates	& & \multicolumn{2}{l}{78°09´N}			\\
  		 	& & \multicolumn{2}{l}{16°02´E}			\\
Geomagn. Inclination	& & \multicolumn{2}{l}{82°06´N}			\\
Invariant Latitude	& & \multicolumn{2}{l}{75°18´N}			\\
Band	                & & \multicolumn{2}{l}{UHF}			\\
Frequency (MHz)         & & \multicolumn{2}{l}{500}			\\
Max. TX bandwith (MHz)  & & \multicolumn{2}{l}{10}			\\
Transmitter 	        & & \multicolumn{3}{l}{16 klystrons}		\\
TX Channels	        & & \multicolumn{3}{l}{Continuously tuneable}	\\
Peak power (MW)         & & \multicolumn{2}{l}{1.0}			\\
Average power (MW)      & & \multicolumn{2}{l}{0.25}			\\
Pulse duration (msec)	& & \multicolumn{2}{l}{$<.001-2.0$}		\\
Phase coding		& & \multicolumn{2}{l}{binary}			\\
Min. interpulse (msec)	& & \multicolumn{2}{l}{0.1}			\\
Receiver		& & \multicolumn{3}{l}{analog-digital}		\\
System temp. (K)	& & \multicolumn{2}{l}{55-65}			\\
Digital processing	& & \multicolumn{3}{l}{12 bit ADC,}		\\
		        & & \multicolumn{3}{l}{lag profiles 32 bit complex}	\\
			& \multicolumn{2}{l}{Antenna 1}		& Antenna 2	\\ \cline{2-4}
Antenna	     		& \multicolumn{2}{l}{dish}		& dish		\\
	     		& \multicolumn{2}{l}{\sq{32}m}		& \sq{42}m Fixed\\
Feed system  		& \multicolumn{2}{l}{Cassegrain}	& Cassegrain	\\
Gain (dBi)   		& \multicolumn{2}{l}{42.5}		& 45		\\
Polarization 		& \multicolumn{2}{l}{circular}		& circular	\\

\end{tabular}

\caption{Table summarising the EISCAT Svalbard radars} \label{esrtable}
\end{table}

The EISCAT mainland radars situated in the Auroral zone and the
ESR in the vicinity of the Cusp and Polar Cap boundary, constitute
an ideal system for the exploration of the Arctic Ionosphere.  The ESR
is the world's most modern IS radar, with capabilities matching or
exceeding all others.  Although some other radars can boast larger
antennas or more powerful transmitters, and hence higher sensitivity, the
ESR with its TV-type transmitter is capable of higher duty cycle than
any other IS radar, which helps to make up for a smaller instrument.  The
modern receiver system has a flexibility and programmability which gives
users a large number of options when it comes to creating their own
experiments.  This is particularly true of the new hardware upgrade of the
mainland radars, which will soon be installed on Svalbard as well.  The
experiment catalogue for the radars contain experiments which cover the
entire Ionosphere from $70$ to \sq{>1200}{km}, with new experiments
under development.

Despite most radar scientists' discussion of the (power) spectra of the
incoherent scattering, the quantity actually measured (or estimated) by
the radar is its Fourier transform equivalent, the autocorrelation
function (ACF) of the scattering through lagged products of samples of the
scattered signal.  The EISCAT radars, instead of forming ACF estimates
from each of a given number of ranges (called gates), sum and store all
such lagged products in a \emph{lag profile matrix}.  Any sufficiently
advanced analysis can then use the information from all lagged products
that contribute to the scattering from a given range to infer the
macroscopic plasma parameters at this range.  When such analysis is done
on an entire ionospheric profile at once, this is called \emph{full
profile analysis} \citep{Holt:1992}.  Although full profile analysis has
been demonstrated \citep{Holt:1992,Lehtinen:1996b}, it is not yet in common
use for IS radar data.

The spectra that are presented in the following section are formed from
the lag profile matrix by summing lagged products in such a way that the
regions contributing to the measurement are roughly equivalent for the
different lags of the autocorrelation function, thus producing a spectrum
from a fairly well-defined and limited region of space.

\subsection{Examples of observations}

\begin{figure}[p]
    \centering\includegraphics[width=5in]
       {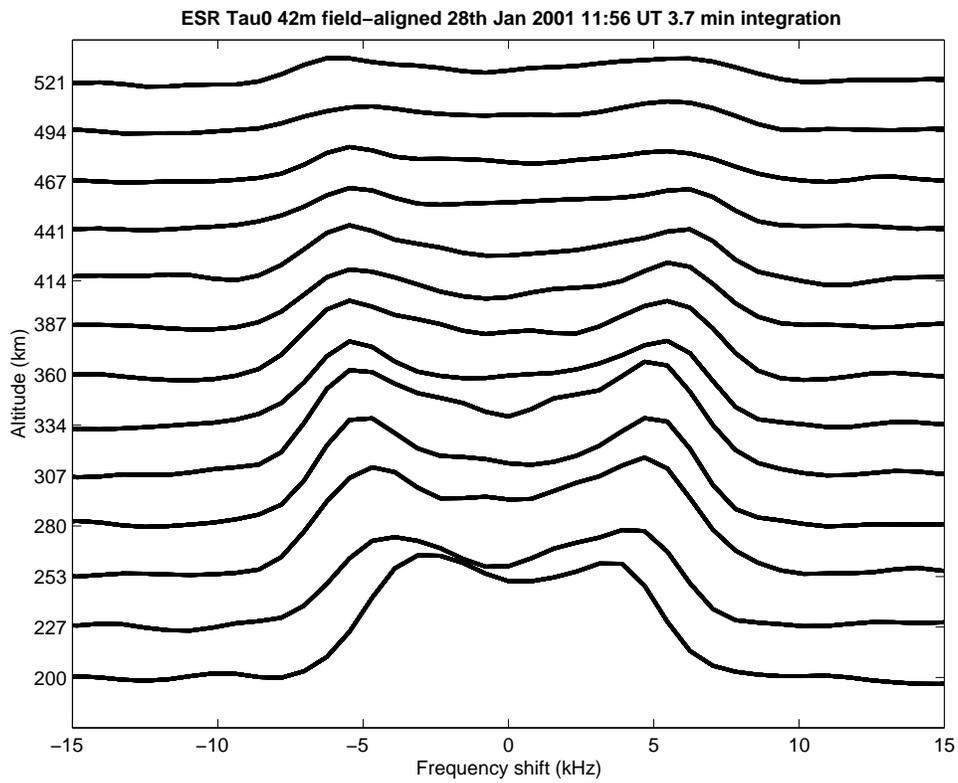}
\caption{Example of spectra taken using the \texttt{tau0} experiment on
the \sq{42}{m} (field-aligned) antenna of the ESR.  These spectra have
been obtained using \sq{3.7}{min} integration.}
  \label{fig:spectra}
\end{figure}

The mainland EISCAT radars were situated in a valley in order to
eliminate the problem of unwanted scattering from the ground, so-called
\emph{ground clutter}.  The close-in mountains scatter only at such short
ranges that the receiver has not yet been opened, while shielding any
solid scatterers at longer ranges.  With its location on a mountain, the
ESR has no such shielding and, for its first few years of operation, the
problem of eliminating ground clutter restricted the lower range of its
observations to approximately \sq{150}{km} beyond the range of the
farthest mountain visible from the site.  During this period, an
experiment consisting of four long pulses (uncoded) called \texttt{gup0}
was used almost exclusively.

With the solution of the ground clutter problem \citep{Turunen:2000}, a
new experiment called \texttt{gup3} was introduced as the standard
experiment in 1999.  This experiment combines long pulses and alternating
codes \citep{Lehtinen:1987}, with the alternating codes covering the $E$
and lower $F$ regions, and the long pulses extending coverage into the
topside Ionosphere.

More recently, an experiment called \texttt{tau0}, using alternating codes
exclusively, has been adopted as the standard experiment at the ESR.
This experiment provides coverage from $90$ to \sq{1100}{km}, and is
typically used with \sq{6.4}{s} time resolution, although $12.8$ and
\sq{3.2}{s} have been used on occasion.  In figure~\ref{fig:spectra},
we show a typical example of \texttt{tau0} spectra at $F$-region
altitudes, using \sq{3.7}{min} integration.  We observe how the width of
the spectra increase with altitude, indicating higher ion temperatures,
and how the scattering power decreases, indicating lower electron
densities.

\subsection{Analysed data}

As discussed above, we can infer a number of macroscopic plasma parameters
from the power spectrum or ACF of the scattering.  This process is called
the \emph{analysis} of the (raw) data, and the result is called
\emph{analysed data}.  The program used for analysis of EISCAT data is
called GUISDAP, or Grand Unified Incoherent Scatter Design and Analysis
Program \citep{Lehtinen:1996a}.

In colour plate 1, we have plotted \sq{96}{hours} of an 18-day continuous
experiment conducted on February 5.--23., 2001.  This experiment
illustrates the reliability of the radar and its capability of obtaining
long time series observations.  In the plate, we can see how the electron
density and temperature decreases during nighttime, and the extreme
variability of the polar ionosphere.  Such data are available from the
online analysed data archive at \texttt{www.eiscat.uit.no}, through the
MADRIGAL database.

\section{Other instrumentation on Svalbard}

The excellent facilities of the ESR are further strengthened by a number
of instruments located in the immediate vicinity of the ESR or with
overlapping fields of view.  Also mentioned here are instruments in the
vicinity of the magnetic conjugate point of the ESR.

The University of Tromsø owns an optical aurora station operated by the
University courses on Svalbard (UNIS) in Adventdalen, outside
Longyearbyen, and many universities and institutions around the world have
their instruments at the station.  In particular the Optics Group at the
Geophysical Institute, University of Alaska contribute both with
instruments and finances.  Amongst the instruments hosted at the station
are various all-sky cameras, a Meridian Scanning Photometer (MSP), an
Auroral Spectrograph, a Michelson Interferometer, several Eber-Fastie
Spectrometers, and magnetometers.  \cite{Sigernes:2001}.
Correspondingly, the Zhongshan station is equipped with all-sky TV cameras
\egcite{Hu:1999}, 

In 1995, an $8\times8$ beam Imaging Riometer for Ionospheric Studies
(IRIS) was installed in Adventdalen, close to the auroral station, with
contributions from the Danish Meteorological Institute (DMI), the National
Institute of Polar Research of Japan (NIPR), UNIS of Longyearbyen and the
University of Tromsø.  Equivalent instruments are also installed at the
South Pole station, at the Antarctic Syowa and Zongshan stations, at
Equaliut in Canada, Søndre Strømfjord and Danmarkshavn on Greenland,
Tjørnes on Iceland, Kilpisjärvi in Finland and outside Ny-Ålesund on
Svalbard \citep{MEL:Stauning}.  The riometer measures the absorption of
cosmic radiation, an absorption usually caused by energetic precipitation
penetrating to the lower Ionosphere.  The instrument has 64 antennas which
are used to form 64 beams covering an area of \sq{240\times240}{km} at
\sq{90}{km} altitude, and the instrument obtains a full image every
second.  A closer description of the instrument and its operations is
given in \citep{Detrick:1990}.

The SOUSY HF radar is situated at the foot of the mountain of the ESR.
It is a phased-array system using 356 Yagi antennas with a 4° wide beam at
the zenith, or 5° zenith angle in either of the NE, NW, SE or SW
directions, and an operating frequency of \sq{53.5}{MHz}
\citep{Rottger:2000}.  Being an MST radar, it is suited for observations
of the stratosphere, mesosphere and thermosphere, and in particular Polar
Mesospheric Summer Echoes (PMSE)

Svalbard is also covered by the CUTLASS pair of the Arctic Dual Auroral
Radar Network (SuperDARN).  The Arctic SuperDARN network covers most of
the Auroral zone (except Siberia), and provides good wide-area convection
pattern coverage around Svalbard \citep{Greenwald:1995}.  In the
Antarctic, another network of SuperDARN radars is under development, with
the Eastward field of view of the Syowa station covering the area around
the Zhongshan and Davis stations \citep{Ogawa:1996}.

The University of Leicester, UK, operated an ionosonde in Longyearbyen
until recently.  The instrument is not operational at the moment, but it
will be moved to the SPEAR (described below) site for reactivation
shortly.

In addition, a heating facility called Space Plasma Exploration by Active
Radar (SPEAR) is under construction beside the ESR \citep{Wright:2000}.
Like the heating facility outside Tromsø, this facility can be used for
ionospheric modification experiments, induced plasma lines, and to produce
$E$-region irregularities which will act as scatterers for the SuperDARN
HF radars.

Ny Ålesund, \sq{\approx150}{km} Northwest (magnetically almost North) of
Longyearbyen is a busy research community, and in addition to the
instruments already mentioned, it hosts a rocket launching facility
(SvalRak) which is used for the launch of ionospheric research
rockets; since the launch site is at 79° North it is ideally located for
scientific exploration of the dayside aurora and processes in the
magnetospheric boundary layer \citep{Maynard:2000}.  The Alfred Wegener
Institute for Polar and Marine Research operates a multiwavelength Lidar
facility at Ny Ålesund, which monitors mainly the middle atmosphere, but
which can link temperature measurements from the ionosphere with those
from the neutral atmosphere at the mesopause \citep{Neuber:1998}.

\section{The University Courses on Svalbard (UNIS)}

The four Universities of Norway have cooperated to establish the
University courses on Svalbard, offering courses and degrees to students
from all of the world in four areas:  Arctic Biology, Arctic Geology,
Arctic Geophysics and Arctic Technology.  Each year, 100 students are
admitted to the undergraduate programmes, and 35 different courses are
given.  The programme in Arctic Geophysics includes a course on the upper
polar atmosphere, and the instruments at the Adventdalen station and the
EISCAT Svalbard Radar is used in this course.

\section{Conjugate studies}

By magnetically conjugate, we usually mean two points of the Earth's
ionosphere that are connected by a magnetic field line.
At the magnetic latitude of Longyearbyen ($\approx$ 75°N), the field
lines will usually be open to the interplanetary magnetic field (IMF) or
close far back in the geomagnetic tail.  For open field lines, conjugate
points are usually taken to be points that would be connected by a field
line in the absence of the IMF.  In either case, no conjugacy
through direct linkage along magnetic field lines should be expected at
such high latitudes.  Rather, coordinated inter-hemispheric studies should
be used to explore the extent to which magnetospheric symmetry is
maintained or breaks down during substorms and auroral displays.
Asymmetries between the nighttime and daytime ionosphere can also be
investigated.

An early systematic conjugate study of visual aurora was a
comparison of all-sky camera images from Alaska and the Antarctica
at $L\approx4$ \citep{DeWitt:1962}.   Later, a comparison of all-sky
camera images from the Antarctic Syowa station and Reykjavik was made,
concluding that the situation was less clear at this higher latitude
$L\approx6$ than at the lower latitude \citep{Wescott:1966}.  The famous
conjugate flights carried out between 1967 and 1974
\citep{Belon:1969,StenbaekNielsen:1972,StenbaekNielsen:1973} resulted in
excellent night-time conjugate auroral all-sky images on a number of
occations, results which are further discussed in
\citep{StenbaekNielsen:1997}.

Riometers have also been used extensively to study conjugate phenomena.
Initially, single wide-beam riometers were used
\egcite{Leinbach:1963,Eriksen:1964},
while more recent studies have used riometers with multiple narrow beams
\egcite{Lambert:1992}, which enables the derivation of velocity vectors
for the motion of absorption regions.  Lately, a series of imaging
riometers have been established in the Southern and Northern auroral
zones \citep{Nishino:1999}, and conjugate observations combining imaging
riometers and TV cameras have been reported \citep{MEL:Yamagishi}.

As the Earth's magnetic field is perturbed by external influences, the
point magnetically conjugate to any given location moves around.
For the Antarctic Zhongshan station, the conjugate point usually lies to
the West of Svalbard, towards Greenland.  \citep{MEL:Yamagishi}.  For
phenomena with larger footprints, like magnetic field disturbances
measured on the ground, it it more appropriate to talk of a conjugate
region.  Conjugate studies should therefore always employ instruments with
a wide field of view (all-sky cameras, imaging riometers, scanning
photometers) or instruments which measure extended phenomena
(e.g.~magnetometers).  Satellite instruments can also help through
large-scale imaging of auroral forms and \emph{in-situ} measurements of
particle fluxes and magnetic field vectors.  Incoherent scatter radars
measure in only one direction at a time, but they supply information on
the physical parameters as a function of range along this direction,
instead of the integrated quantities available through most of the other
instruments discussed here.  This can provide details on the entire energy
spectrum of precipitating particles otherwise unattainable from the
ground.  With real-time determination of conjugacy from optical or imaging
riometer observations, the radar can be pointed in the direction of the
conjugate region for pinpointed observations.  ISR is also capable of
operating during daylight or overcast conditions (unlike optical
instruments), and of providing continuous coverage over long periods of
time (unlike satellite-borne instruments and rockets).
The extensive and detailed information derived from the incoherent scatter
technique, coupled with the coherent radar and optical data available on
Svalbard represents a large untapped opportunity for effective conjugate
studies in co-operation with Chinese scientists to further investigate
detailed differences between the Northern and Southern Polar regions.

\section{Conclusion}

We have described in some detail the EISCAT Svalbard Radar (ESR) and some
of the instruments located in its vicinity.  Through a review of
previous conjugate studies, we have shown how this capable instrument and
the instrument cluster can contribute to interhemispheric coordinated
observations.

To date, the ESR has not participated in any conjugate point studies but
extensive datasets are already available and dedicated observation
programs can be scheduled in the future.

\section{Acknowledgements}

The EISCAT Scientific Association is supported by the \emph{Centre
National de la Recherche Scientifique} of France, the
\emph{Max-Planck-Gesellschaft} of Germany, the {Particle Physics and
Astronomy Research Council} of the United Kingdom, \emph{Norges
Forskningsråd} of Norway, \emph{Naturvetenskapliga Forskningsrådet} of
Sweden, \emph{Suomen Akatemia} of Finland and the \emph{National Institute
of Polar Research} of Japan.

Two of the authors (TG and AS) are supported through grants from the NFR
of Norway.

\bibliography{refs}

\end{document}